\documentclass{aa}
\usepackage{graphicx}
\usepackage{times}
\newcommand{\etal}{{et al.}}

\newcommand{\beq}{\begin{equation}}
\newcommand{\eeq}{\end{equation}}

\begin{document}

%==============================================================================
\title{Line formation in convective stellar atmospheres}
\subtitle{I.Granulation corrections for solar photospheric abundances}

\author{
Matthias Steffen\inst{1}
\and
Hartmut Holweger\inst{2}}
\offprints{Matthias Steffen}
\institute{
  Astrophysikalisches Institut Potsdam,
  An der Sternwarte 16,
  14482 Potsdam, Germany
  [MSteffen@aip.de]
\and
  Institut f\"ur Theoretische Physik und Astrophysik,
  Universit\"at Kiel,
  24098 Kiel, Germany
  [holweger@astrophysik.uni-kiel.de]
}
\date{Received 31-10-01; accepted 01-03-02}

%==============================================================================
\abstract{
In an effort to estimate the largely unknown effects of photospheric
temperature fluctuations on spectroscopic abundance determinations, we
have studied the problem of \emph{LTE line formation in the
inhomogeneous solar photosphere} based on detailed 2-dimensional
radiation hydrodynamics simulations of the convective surface layers
of the Sun.  By means of a strictly differential 1D/2D comparison of
the emergent equivalent widths, we have derived \emph{``granulation
abundance corrections''} for individual lines, which have to be
applied to standard abundance determinations based on homogeneous 1D
model atmospheres in order to correct for the influence of the
photospheric temperature fluctuations. 
In general, we find a \emph{line strengthening}
in the presence of temperature inhomogeneities as a consequence of the
non-linear temperature dependence of the line opacity.
The resulting corrections are negligible for lines with an excitation
potential around $E_i=5$~eV, regardless of element and ionization
stage. Moderate granulation effects ($\Delta_{\rm gran} \approx
-0.1$~dex) are obtained for weak, high-excitation lines ($E_i \ga
10$~eV) of C\,{\sc i}, N\,{\sc i}, O\,{\sc i} as well as Mg\,{\sc ii} 
and Si\,{\sc ii}. The largest corrections are found for ground state 
lines ($E_i=0$~eV) of neutral atoms with an ionization potential between
6 and 8~eV like Mg\,{\sc i}, Ca\,{\sc i}, Ti\,{\sc i}, Fe\,{\sc i}, 
amounting to $\Delta_{\rm gran} \approx -0.3$~dex in the case of Ti\,{\sc i}.
For many lines of practical relevance, the magnitude of the abundance
correction may be estimated from interpolation in the tables and graphs 
provided with this paper. The application of
abundance corrections may often be an acceptable alternative to a
detailed fitting of individual line profiles based on hydrodynamical
simulations. The present study should be helpful in providing upper
bounds for possible errors of spectroscopic abundance analyses, and
for identifying spectral lines which are least sensitive to the
influence of photospheric temperature inhomogeneities.
}

\maketitle
\unboldmath

%==============================================================================
\section{Introduction}
\label{S1}
%==============================================================================
Theoretical model atmospheres serve as the fundamental tool for the
quantitative analysis of stellar spectra.  The basic parameters of
a stellar atmosphere -- effective temperature, surface gravity, and
chemical composition -- can be inferred from the comparison between
observed and calculated spectra.  Much effort has been spent to
improve the theoretical models by incorporating as much realistic
physics as possible.  Specifically, in late-type stars one is
confronted with the problem of a detailed treatment of both radiative
and convective energy transport.  A high degree of sophistication has
been achieved in modeling radiative transfer: standard stellar
atmosphere codes can now handle the influence of millions of
spectral lines.

In contrast, the treatment of convection is still rather crude.
Usually, stellar atmospheres are described as 1-dimensional
hydrostatic configurations where the convective energy transport is
calculated from the so-called mixing-length theory (MLT, Vitense
\cite{V53}; B\"ohm-Vitense \cite{BV58}) or variants thereof (e.g.\
Canuto \etal \cite{CGM96}). However, modern spectroscopic
observations have reached a level of precision that requires improved
model atmospheres for a reliable interpretation, including a proper 
treatment of hydrodynamical phenomena.

From the point of view of standard model atmospheres, convection
affects the \emph{temperature structure} in a twofold way: it
influences the \emph{mean vertical stratification} and introduces
\emph{horizontal inhomogeneities}. MLT is designed to model only
the mean structure and is not suitable to construct realistic
multi-component model atmospheres. Although realistic
hydrodynamical model atmospheres do exist for the Sun and a few
solar-type stars due to the pioneering work by Nordlund (\cite{No82}),
Nordlund \& Dravins (\cite{ND90}), Stein \& Nordlund (\cite{SN89},
\cite{SN98}), and Ludwig et al. (\cite{LFS99}), to name just a few
important contributions, such complex models are not usually applied
for the analysis of stellar spectra.

Observationally, the solar granulation is the visible imprint of
thermal convection extending far into the photospheric layers where
spectral lines are formed.  The effect of the associated photospheric
temperature inhomogeneities on the line formation process and the
consequences for spectroscopic abundance determinations are not yet
well understood, but it is generally believed that the errors
introduced by representing a dynamic, inhomogeneous atmosphere by a
static, plane-parallel model are small. However, reliable
quantitative estimates of this effect are difficult to find. Earlier
studies suffer from uncertainties in the underlying empirical 
two-component models, which are rather \emph{ad hoc} and lack a firm 
physical foundation (e.g. Hermsen \cite{He82}). Later investigations
based on relatively crude numerical convection models (Atroshchenko \&
Gadun \cite{AG94}; Gadun \& Pavlenko \cite{GP97}) led to questionable
and partly ambiguous conclusions. Recent determinations of the solar
photospheric Fe and Si abundance by Asplund et al.\ (\cite{ANTS00})
and Asplund (\cite{As00a}), respectively, rely on state-of-the-art 3D
numerical convection simulations and take all hydrodynamical
effects fully into account.  However, comparison of these results with
those based on the empirical 1D Holweger-M\"uller (\cite{HM74}) solar
atmosphere does not allow to separate the influence of inhomogeneities
from the effect of the mean stratification, and to study the 1D/3D
abundance corrections as a function of the line parameters in a
systematic way.

In this work we investigate the \emph{spectroscopic effects of
horizontal temperature inhomogeneities in the solar atmosphere} based
on detailed 2D hydrodynamical models of solar surface convection
(Freytag \etal\ \cite{FLS96}; Ludwig \etal\ \cite{LFS99}, Steffen
\cite{S00}). We do \emph{not} address the question whether the mean 
temperature structure supplied by MLT is an appropriate representation
of the mean stratification obtained from hydrodynamical simulations 
(the problem of the `right' choice of the mixing-length parameter 
$\alpha$ for recovering the correct mean temperature
stratification was investigated by Steffen \etal\ (\cite{SLF95})
in the context of white dwarf atmospheres). 
Rather, our main question is: how large are the systematic errors 
of standard spectroscopic abundance determinations due to the fact that
one replaces the `real' line profile, which actually is the result of
averaging the spatially resolved line \emph{intensities} over the granulation
pattern, by the line profile computed from a single plane-parallel
temperature stratification?
 
In Sect.~\ref{S2}, we introduce a simple analytical argument to
illustrate the basic effect of line strengthening in the presence of
temperature fluctuations. Then we summarize the main characteristics
of the numerical simulations of solar surface convection in
Sect.~\ref{S3}, outline our method to derive `granulation abundance
corrections' from a differential 1D/2D comparison of computed line
equivalent widths in Sect.~\ref{S4}, before finally presenting the
results for the Sun in Sect.~\ref{S5}.  A summary of our findings is
given in Sect.~\ref{concl}.

In order to give a physical explanation for the numerical results
obtained in this work, we have carried out a detailed investigation
the process of LTE line formation in an inhomogeneous stellar
atmosphere based on the analysis of the transfer equation and the
evaluation of line depression contribution functions. This study is
presented in the second paper of this series (Paper~II).
 
%==============================================================================
\section{A simple analytical illustration}
\label{S2}
%==============================================================================
The strength of a spectral line formed in a plane-parallel (1D)
atmosphere will differ from the strength of the same line formed in an
inhomogeneous atmosphere with the same chemical composition and 
mean temperature stratification
because, in general, there is a \emph{non-linear relationship between line
strength and temperature}.  It may be expected that lines with a very high
excitation potential are most affected because (i) they
originate from deep layers where convection is well developed, and
(ii) they are highly sensitive to temperature fluctuations. 
Likewise, the molecular equilibrium is known to be
highly sensitive to temperature fluctuations, so molecular lines might
also be strongly affected if the temperature inhomogeneities reach the
higher photosphere where these lines are formed.

% Plot: Fig0
\begin{figure}
\resizebox{\hsize}{!}
  {\mbox{\includegraphics[bb=56 56 538 350]{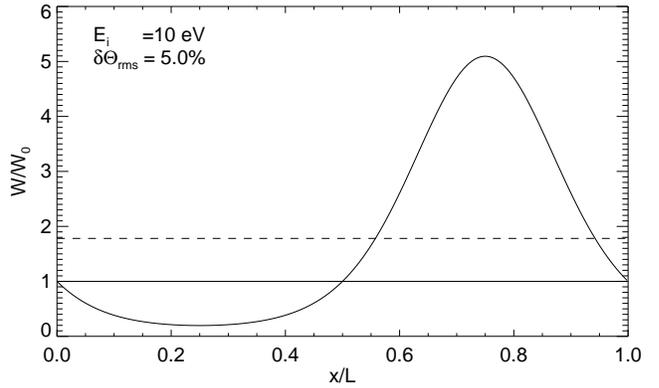}}}
  \caption[]{%
   Spatial variation of $W$ for a sinusoidal temperature variation
   according to Eq.(\ref{E2.6}), for $\sigma=0.05$ and $E_i=10$~eV. 
   Averaging over one spatial wavelength gives 
   $\langle W \rangle/W_0 = 1.78$ (dotted horizontal line).
} % end of caption
  \label{F00}
\end{figure}
%======================================================================
As an illustration of the expected effect, consider the following simple
analytical argument. Assume that the equivalent width of a weak line $W$
depends exponentially on the inverse temperature $\theta=5040/T$, 
as suggested by the form of the Saha-Boltzmann equations,
\footnote{Note that Eq.(\ref{E2.1}) is a reasonable approximation 
only for weak lines of `majority species' with high excitation
potential (cf.\ Gray \cite{Gr92}).}
\beq
W(\Delta\theta)=W_0\exp\left\{-E_i \Delta\theta \ln 10\right\},
\label{E2.1}
\eeq
where $E_i$ is the excitation potential in eV and 
$\Delta\theta = \theta - \theta_0$.
Assume further that $\theta$ is normally distributed 
about the mean $\theta_0$ with standard deviation $\sigma$, 
\beq
g(\Delta\theta)=\frac{1}{\sigma \sqrt{2 \pi}} 
\exp \left\{-\frac{\Delta\theta^2}{2 \sigma^2}\right\}.
\label{E2.2}
\eeq
Then it can be shown by analytical integration that
\begin{eqnarray}
\langle W\rangle&=&\int_{-\infty}^{+\infty}
g(\Delta\theta)\, W(\Delta\theta)\, \mathrm{d}\Delta\theta \nonumber \\ 
&=&W_0 \exp\left\{\frac{1}{2}E_i^2 \sigma^2 (\ln 10)^2\right\}
\label{E2.3}
\end{eqnarray}
i.e.\ the mean value $\langle W \rangle$ of the equivalent width
formed in an inhomogeneous medium is always greater than the
equivalent width for the unperturbed medium, $W_0$. As long as $E_i
\sigma$ is small enough, the effect is negligible, but this is not
true in general.  Assuming $\sigma = 0.05$, corresponding to 6\%
temperature fluctuation around $T_0$ = 6000~K, and $E_i = 10$~eV, the
effect is very substantial: $\langle W\rangle \approx 2\,W_0$, i.e.
the elemental abundance derived from such lines would be
\emph{overestimated} by a factor of 2! Note that, for example, several
important neutral oxygen lines have $E_i$ between 9 and 11 eV.

A similar result is obtained when assuming a totally different distribution
of $\Delta\theta$, 
\beq
f(\Delta\theta) = \frac{1}{\pi \sqrt{2 \sigma^2-\Delta\theta^2}},
\label{E2.4}
\eeq
corresponding to sinusoidal temperature fluctuations in one horizontal
direction,
\beq
\Delta\theta = \sigma\sqrt{2}\sin(2\pi x),
\label{E2.5}
\eeq
where $x$ is the spatial coordinate. According to Eq.(\ref{E2.1}),
the fluctuation of the equivalent width is then 
\beq
W(x)=W_0\exp\left\{-E_i\ln 10\,\sigma\sqrt{2}\sin(2\pi x)\right\}.
\label{E2.6}
\eeq
Adopting again $\sigma = 0.05$ and $E_i = 10$~eV, we have plotted
$W(x)$ in Fig.~\ref{F00}. Clearly, the fluctuation of $W$ is highly
asymmetric with respect to $W_0$, and numerical averaging over one spatial 
wavelength gives $\langle W \rangle \approx 1.8 \,W_0$.

Looking at Table~\ref{tab1}, it turns out that the
simple model severely overestimates the abundance correction in the
case of the oxygen lines mentioned as an example above. This is
because, in reality, the line formation process is much more
complicated, since the line strength depends not only on the line
opacity but also on the continuum opacity and on the local temperature
gradient of the atmosphere (see Paper~II).  Moreover, spectral lines
usually form over a substantial depth range, and the sign of $\Delta
\theta$ may depend on depth, leading to partial cancellation of the
local effects, especially for stronger lines. Finally, saturation can play
a significant role as well. Hence, realistic abundance corrections can
only be derived from detailed line formation calculations based on
multi-dimensional hydrodynamical model atmospheres.

%======================================================================
% Plot: Fig1
\begin{figure*}
\resizebox{\hsize}{!}
  {\mbox{\includegraphics[bb=20 60 250 800, angle=90,width=17cm,clip=true] 
{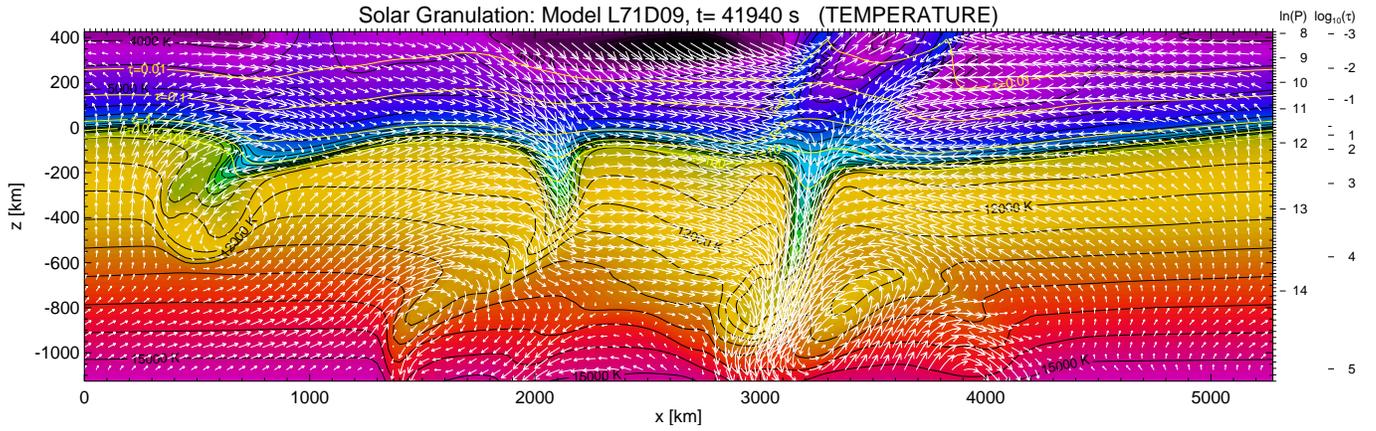}}}
  \caption[]{%
  Snapshot from a \mbox{2-dimensional} numerical
  simulation of solar surface convection after 41\,940~s of simulated
  time. This model was computed on a Cartesian grid with 210x106 mesh
  points (tick marks along upper and right side), with periodic
  lateral boundary conditions ($L = 5\,250$~km). ``Open'' boundary
  conditions at the bottom and top of the computational domain are
  designed to minimize artificial distortions of the flow. The
  velocity field is represented by pseudo streamlines, indicating the
  displacement of a test particle over 20~sec (maximum velocity is
  10.1~km/s at this moment); the temperature structure is outlined by
  temperature contours in steps of 500~K. Geometrical height $z=0$
  (scale at left) corresponds to $\tau_{\rm Ross} \approx 1$; scales
  at right refer to the horizontally averaged gas pressure \mbox{[dyn
  cm$^{-2}$]} and Rosseland optical depth $\tau_{\rm Ross}$.
} % end of caption
  \label{L71D09_625}
\end{figure*}
%======================================================================
%==============================================================================
\section{Time-dependent Radiation Hydrodynamics}
\label{S3}
%==============================================================================
In order to adequately address the complex problem outlined above, we
have investigated the effects of temperature inhomogeneities on the
formation of a variety of spectral lines by means of a realistic 2-D
radiation hydrodynamics (RHD) simulation of solar surface
convection. In the following, we briefly summarize the foundations of
the hydrodynamical simulation and the main results concerning the
solar photosphere.

\subsection{2-dimensional numerical simulation}
Stellar surface convection is governed by the conservation equations of 
hydrodynamics, coupled with the equations of radiative energy transfer.
Our hydrodynamical models result from the numerical integration of
this set of partial differential equations. This approach constitutes 
an increasingly powerful tool to study in detail the 
time-dependent hydrodynamical properties of the solar granulation. 

Just like `classical' stellar atmospheres, the hydrodynamical models
are characterized by effective temperature, $T_{\rm eff}$, surface
gravity, $g$, and chemical composition of the stellar matter.  But in
contrast to the mixing-length models, they account for `overshoot' and
no longer have any free parameter to adjust the efficiency of the
convective energy transport.  Based on first principles, RHD
simulations provide physically consistent \emph{ab initio} models of
stellar convection which can serve to address a variety of questions,
including the problem of line formation in inhomogeneous stellar
atmospheres.

The models used for the present investigation comprise a small section
near the solar surface, extending over 7 pressure scale heights in the
vertical direction, including the photosphere, the thermal boundary
layer near optical depth $\tau_{\rm Ross}=1$, and parts of the
subphotospheric layers.  Only the uppermost layers of the deep solar
convection zone can be included in the model, requiring an open
lower boundary.
The simulations are designed to resolve the solar granulation. 
The effect of the smaller scales, which cannot be resolved numerically, 
is modeled by means of a subgrid scale viscosity (so-called Large Eddy 
Simulation approach). Spatial scales larger than the computational box
are unaccounted for. We employ a realistic equation of state,
including ionization of H, HeI, HeII and H$_2$ molecule formation. 
In order to avoid problematic simplifications like the diffusion or
Eddington approximation, we solve the non-local radiative transfer
problem along a large number 
($N_{\rm x}\,f_{\rm sub}\,N_\theta\,N_\phi 
= 210 \cdot 4 \cdot 2 \cdot 8 \approx 10\,000$) of rays in 5
\emph{opacity bins}, with realistic opacities accounting adequately for
the influence of spectral lines (see Nordlund \cite{No82}, 
Ludwig \cite{Lu92}). For further details see
Ludwig \etal\ (\cite{LJS94}), Freytag \etal\ (\cite{FLS96}), 
Ludwig \etal\ (\cite{LFS99}), Steffen (\cite{S00}).

\subsection{Main results}
\begin{figure}
\resizebox{\hsize}{!}
{\mbox{\includegraphics[bb=28 36 538 376]{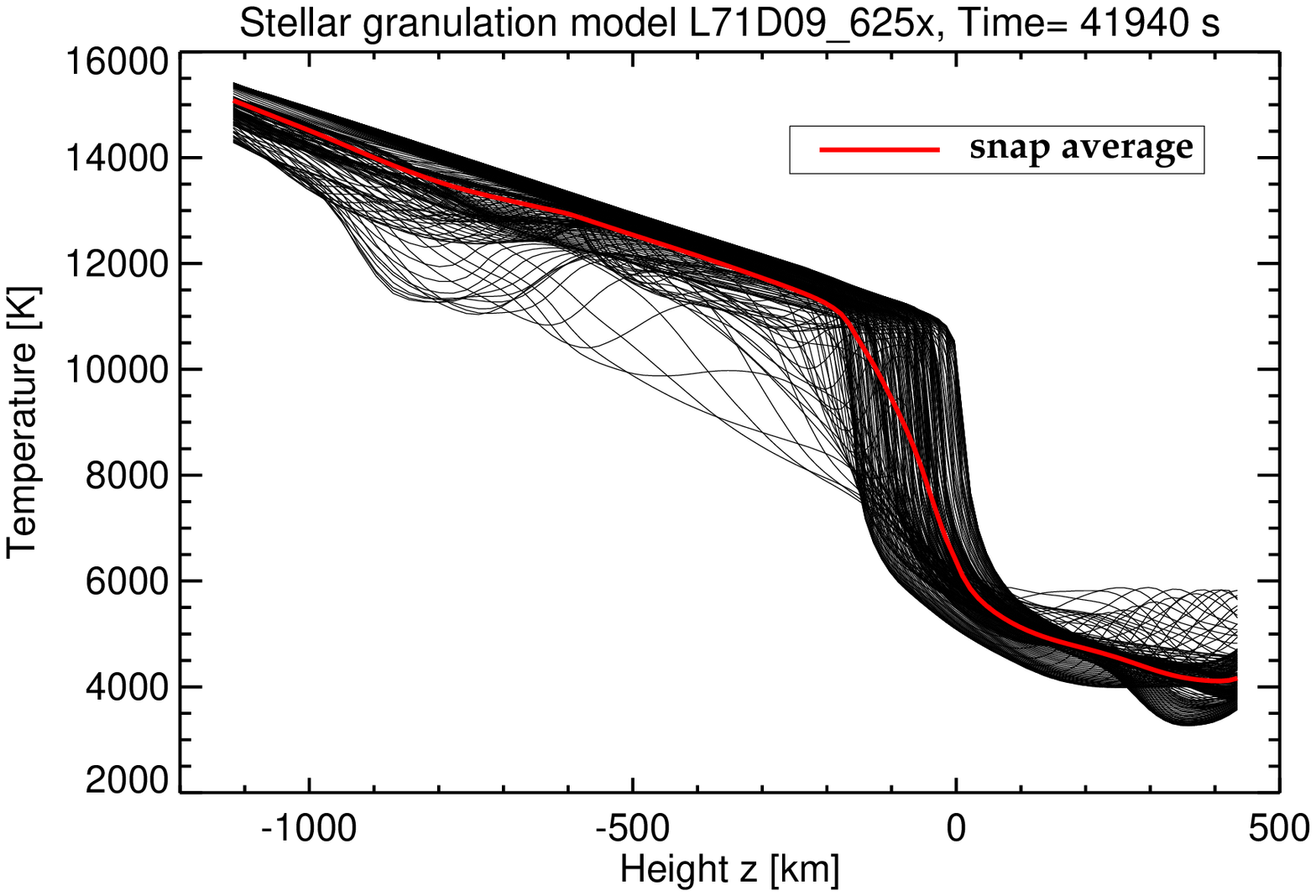}}}\\
\resizebox{\hsize}{!}
{\mbox{\includegraphics[bb=28 36 538 376]{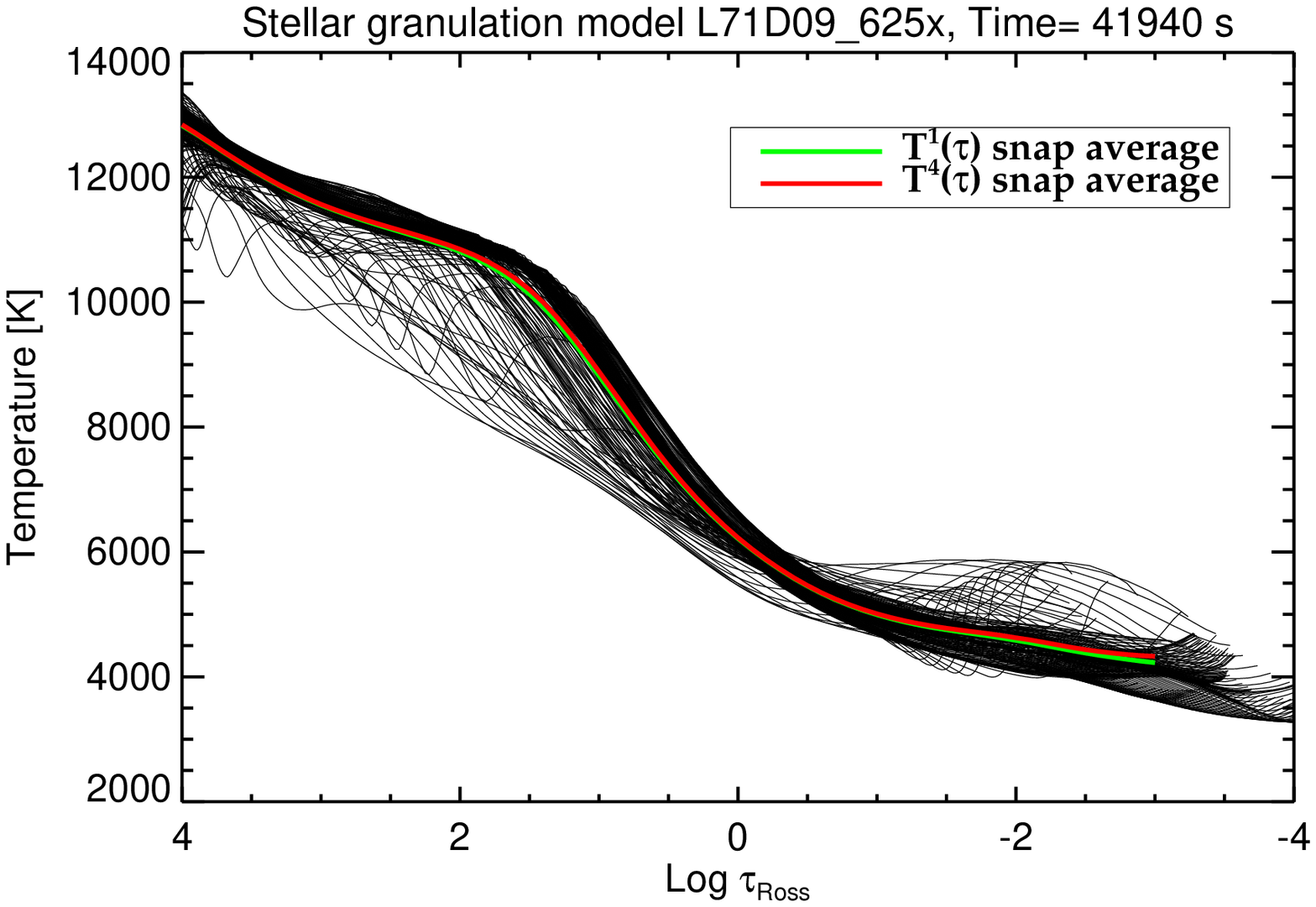}}}
\caption[]{%
Temperature as a function of geometrical depth $z$ (top) and
 as a function of individual optical depth $\tau_{\rm Ross}$ (bottom)
for the snapshot shown in Fig.~\ref{L71D09_625}. Each line corresponds
to a different $x$ position. Note that the $\langle T \rangle$ and the  
$\langle T^4 \rangle^{1/4}$ average are hardly distinguishable in this
representation.
} % end of caption
\label{Tz}
\end{figure}

The basic features of the numerical convection model are illustrated
in Fig.~\ref{L71D09_625}, showing a representative snapshot from a
well-relaxed simulation of the solar granulation.  In the convectively
unstable, subphotospheric layers, fast ($v \la c_s)$ narrow downdrafts
(also called plumes or jets) stand out as the most prominent
feature. They are embedded in broad ascending regions (the granules)
where velocities are significantly lower. We find enormous temperature
differences between the cool downflows and the hot upflows ($\Delta
T_{\rm max} \approx 5\,000$~K at $z \approx -100$~km). 
Near optical depth $\tau=1$, efficient radiative surface cooling
produces a very thin thermal boundary layer over the ascending parts
of the flow, where the temperature drops sharply with height ($\Delta
T/\Delta z \approx 80$~K/km, see upper panel of Fig.~\ref{Tz}) and the
gas density exhibits a local inversion. Note however, that the
temperature contrast is much reduced on the optical depth scale (see
bottom panel of Fig.~\ref{Tz}).

%=========================================================================
% Plot: Fig2
%=========================================================================
Although convectively stable according to the Schwarzschild criterion,
the photosphere ($\tau \la 1$) is by no means static. Convective flows
overshooting into the stable layers from below are decelerated here
and deflected sideways. This can sometimes result in transonic
horizontal streams which lead to the formation of shocks in the
vicinity of strong downdrafts (see vertical fronts in
Fig.~\ref{L71D09_625} between $3\,000 < x <3\,500$~km, $z \ga 100$~km).

Oscillations excited by the stochastic convective motions contribute
to the photospheric velocity field as well. Interestingly, the
oscillation periods lie in the range 150 to 500~s for the solar
simulation, in close agreement with the observed 5 minute
oscillations.  Controlled by the balance between dynamical cooling
due to adiabatic expansion and radiative heating in the spectral
lines, the mean photospheric temperature stratification is slightly 
cooler than in radiative equilibrium, and the resulting temperature 
structure is not at all plane-parallel.

\begin{figure}
\resizebox{\hsize}{!}
{\mbox{\includegraphics[bb=56 28 525 375]{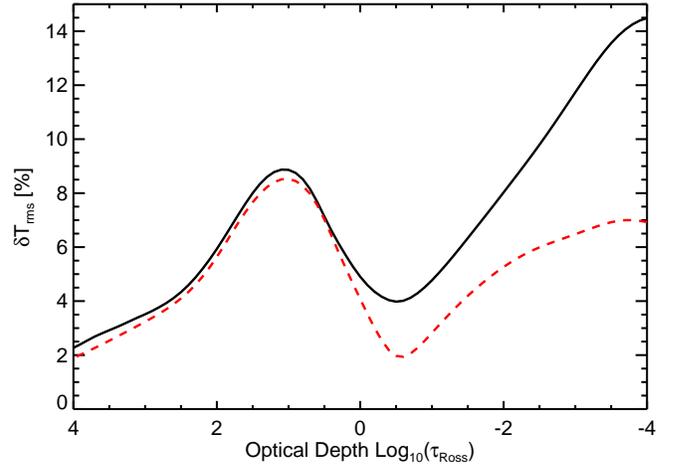}}}
\caption[]{%
Time-averaged amplitude of rms temperature fluctuations as a function
of Rosseland optical depth, derived from the numerical simulation of solar
granulation described in Sect.~\ref{S3} (solid). $\delta T_{\rm rms}$
was obtained by averaging over surfaces of constant Rosseland optical 
depth for each snapshot and subsequently taking a long-term time average.
For comparison, we show corresponding results from a 3-dimensional
numerical simulations (dashed) recently carried out with a new code by 
B.~Freytag and M.~Steffen. 
} % end of caption
\label{Trms}
\end{figure}

According to Eq.(\ref{E2.3}), the line strengthening depends on the key
quantity $\sigma = \theta_0\,\delta T_{\rm rms}/T_0$, where $\delta T_{\rm
rms}=\sqrt{\langle(T-T_0)^2 \rangle}$.  We have evaluated $\delta
T_{\rm rms}$ as a function of optical depth for our 2D numerical
simulation of solar convection (Fig.~\ref{Trms}).  For the problem
under investigation, temperature differences have to be taken over
levels of constant optical depth, rather than at constant geometrical
depth. According to the simulations, typical values in the lower
photosphere are in the range $0.04 < \delta T_{\rm rms} < 0.06$. The
minimum near $\log \tau_{\rm Ross}=-0.5$ is related to the `inversion' of
the temperature fluctuations at this height.

It is worth mentioning that we have recently carried out fully
3-dimensional simulations with a new code developed by B.~Freytag
and M.~Steffen (Freytag; Steffen \etal, in preparation). 
We find that the depth dependence of $\delta T_{\rm rms}(\tau)$ is 
qualitatively similar in 2D and 3D. However, the amplitude of 
the temperature fluctuations is systematically smaller at all heights
in the 3D simulation. The difference is most pronounced in the higher 
photosphere (see Fig.~\ref{Trms}).

\subsection{Accuracy of 2D and 3D convection models}
 The straightforward application of hydrodynamics and 
 radiative transfer simulations is certainly the most
 satisfactory and consistent method to treat the line formation in a
 convective stellar atmosphere, although the outcome is difficult to
 check by independent analysis. The hydrodynamical models
 presented here must be regarded as somewhat preliminary for 
 several reasons.
 First of all, the restriction of the flow geometry to 2 dimensions
 must lead to deviations from the true 3D atmospheric structure. As
 a consequence, the convective line shift and asymmetry predicted by 
 the 2D granulation models is quantitatively different from that
 obtained with 3D models (Asplund et al. \cite{ALNS00}).
 A further problem is the relatively poor spatial resolution of the 
 simulations (especially in the vertical direction), which is barely 
 sufficient to resolve the thin thermal boundary layer near $\tau=1$.
 Certainly, the Reynolds number of the simulations is many orders of
 magnitude below realistic values.
 Finally, the influence of the spectral lines on the radiative energy balance
 can be treated only in an approximate way. Typically, a few opacity bins 
 are used in the simulations, as compared to thousands of frequency points 
 in standard 1D models.

 The latter problems apply to state-of-the-art 3D convection
 simulations as well. We note that a perfect line profile fitting does
 not necessarily imply a perfect model atmosphere. Another important
 diagnostic is the center-to-limb variation of the continuum
 intensity. Even the best 3D solar granulation models still fail to
 reproduce the observed continuum properties with high accuracy
 (Asplund et al.\ \cite {ANT99}), indicating that ``the temperature
 structure close to the continuum forming layers could be somewhat too
 steep''.

 In view of the above mentioned limitations, we prefer to avoid a
 direct application of our convection simulations. Instead, we favor
 the somewhat less powerful but more reliable differential approach
 described in the following section.

%=========================================================================
% Plot: Fig3
%=========================================================================

%==============================================================================
\section{Granulation abundance corrections}
\label{S4}
%==============================================================================
\subsection{The problem}
\label{S41}
The problem we want to address is the following. Assume you
have the `best possible' 1D representation of the solar atmosphere,
correctly reproducing the continuum intensity distribution
$I_\lambda(\lambda)$ and its center-to-limb variation
$I_\lambda(\mu)$. The microturbulence parameter $\xi_{\rm mic}$ is
adjusted such as to minimize the overall dependence of the derived
abundances on line strength.  The empirical Holweger-M\"uller
(\cite{HM74}) atmosphere comes close to this ideal mean temperature
structure. In the presence of temperature fluctuations, however,
non-linearities in the line formation process will inevitably lead 
to systematic errors of abundance determinations based on individual 
spectral lines. The present work aims at quantifying this kind of error.

We have adopted a strictly differential procedure to quantify these
errors.  The basic idea is to identify our 2D hydrodynamical
convection simulations with the real inhomogeneous atmosphere, and the
mean structure of the simulations with the `best possible' 1D
representation. Note that it is not important for the hydrodynamical
convection models to reproduce the mean temperature structure of the
real solar atmosphere with great precision, because the related errors
cancel to a first approximation in our differential
comparison. Similarly, the details of the hydrodynamical velocity
field are of secondary importance in the present context, since such
1D/2D differences are largely eliminated by a proper choice of the 
microturbulence parameter. In contrast, there is no free parameter to
account for the presence photospheric \emph{temperature
inhomogeneities}. This is why we focus on the abundance corrections
related to temperature fluctuations $\delta T_{\rm rms}$ only.
The most important quantity to be provided by the convection model is 
therefore the magnitude and height dependence of $\delta T_{\rm rms}$.

In the following, we outline our method to derive LTE 
``granulation abundance corrections'', $\Delta_{\rm gran}$, which must
be added to the logarithmic elemental abundance derived from standard
1D model atmosphere analysis in order to compensate for the effects of
temperature inhomogeneities.  Note that our procedure ensures that the
corrections $\Delta_{\rm gran} \rightarrow 0$ for vanishing horizontal 
\emph{temperature} and \emph{pressure fluctuations}.

\subsection{The procedure}
\label{S42}
First we have selected $S=11$ representative snapshots from our
time-dependent convection simulation (see Sect.~\ref{S3}). These
snapshots (one of them shown in Figs.~\ref{L71D09_625} and \ref{Tz})
all have the correct effective temperature (the horizontally
averaged radiative energy flux lies within $\pm1\%$ of the nominal
solar flux), but differ in the number and size distribution of
granules and in the continuum intensity contrast ($\delta I_{\rm rms}$
varies between 18.0 and 26.5\% at $\lambda\,5000$ \AA). For each of
the 2D snapshots ($s=1 \, .. \, 11$) we then constructed a \emph{corresponding
1D model} by averaging temperature $T$ and gas pressure $P$ over the corrugated
surfaces of constant Rosseland optical depth:

\beq
T_{\rm s,1D}(\tau_{\rm Ross}) = 
        \left\langle T^4_{\rm s,2D}(\tau_{\rm Ross},x)\right\rangle^{1/4},
\label{T0}
\eeq

\beq
P_{\rm s,1D}(\tau_{\rm Ross}) = 
        \left\langle P_{\rm s,2D}(\tau_{\rm Ross},x)\right\rangle,
\label{P0}
\eeq
where $\langle.\rangle$ denotes averaging over the horizontal coordinate $x$.
Note that the 1D models constructed in this way have very nearly the same
radiative flux as the underlying 2D models. 

Next we compute, for given elemental abundance and atomic line
parameters, the equivalent width of any given spectral line from
(i) the set of 2D snapshots by horizontal averaging of the spatially
resolved synthetic line profile

\beq
W_{\rm s,2D}=\left\langle I_{\rm c,s}(x)\,W_{\rm s,2D}(x)\right\rangle /
\left\langle I_{\rm c,s}(x)\right\rangle, 
\eeq
where $I_{\rm c}(x)$ is the local continuum intensity,
and (ii) from the corresponding 1D models, resulting in $W_{\rm s,1D}$.

For the line formation calculations, we employed the Kiel spectrum
synthesis code LINFOR under the assumption of LTE. For the 1D
atmospheres, we have always adopted a depth-independent
microturbulence velocity of $\xi_{\rm mic}=1$~km/s. For snapshots from
the convection simulations we have computed synthetic spectra for two
different assumptions concerning the non-thermal velocity field:
in case (A) we used the depth-dependent hydrodynamical velocity field
$v(x,z)$ obtained from the simulations, in case (B) we replaced the
hydrodynamical velocity field by a depth-independent microturbulence
velocity $\xi_{\rm mic}=1$~km/s.

As the final step, we calculate the mean equivalent widths 
resulting from the set of 1D and 2D line profiles,
$W_{\rm 1D}=\frac{1}{S}\,\sum_{s=1}^S W_{\rm s,1D}$ and
$W_{\rm 2D}=\frac{1}{S}\,\sum_{s=1}^S W_{\rm s,2D}$, respectively.
For truly weak lines, the (logarithmic) ``granulation abundance correction'' 
would simply be 

\beq
\Delta_{\rm gran}^{\rm weak\;lines} = \log W_{\rm 1D} - \log W{\rm 2D}.
\eeq

In general, however, the equivalent width is not proportional to the
elemental abundance due to saturation. We account for these
curve-of-growth effects in the following way. Knowing that the `true'
abundance, $N_{\rm 2D}$ produces the equivalent width $W_{\rm 2D}$
with the 2D inhomogeneous model atmosphere (and $W_{\rm 1D}$ with the
1D mean structure), we iteratively determine the elemental abundance 
$N_{\rm 1D}$ which produces the equivalent width $W_{\rm 2D}$ with 
the 1D mean stratification. The ``granulation abundance correction'' 
is then given by
\beq
\Delta_{\rm gran} = \log N_{\rm 2D} - \log N_{\rm 1D}.
\eeq
For simplicity, we used the standard HM model atmosphere (Holweger \&
M\"uller \cite{HM74}; $\xi_{\rm mic}=1$~km/s) for this final step of
the procedure ($W_{\rm 1D} \rightarrow N_{\rm 2D}$, $W_{\rm 2D}
\rightarrow N_{\rm 1D}$). This is justified because all we need is a
realistic curve-of-growth, $W(N)$, which is very similar for the
averaged simulation snapshots and the HM atmosphere.

Case (A) and (B) give essentially the same corrections for weak lines,
but the results may differ substantially for stronger lines, where
$W_{\rm 2D,A} < W_{\rm 2D,B}$ (implying that the hydrodynamical
velocity field provides an effective `microturbulence' $\xi_{\rm
mic}({\rm RHD}) < 1$~km/s). For lines of arbitrary strength, we strongly
favor the granulation corrections derived from case (B), because this
strictly differential 1D/2D comparison is based on the same
non-thermal velocity field, and thus cleanly separates the effect of the
\emph{temperature inhomogeneities} from that of the small-scale
velocity field. In the following, $\Delta_{\rm gran}$ therefore
denotes the granulation abundance corrections derived from case (B).

For practical reasons, we have used only a small number of snapshots
($S=11$) for our present investigation. The statistical uncertainty of
the derived abundance corrections is therefore appreciable. In order
to quantify this uncertainty, we have also tabulated the standard
deviation of the corrections $\Delta_{\rm gran}$,
\beq
\sigma = \sqrt{\frac{1}{S-1}\sum_{s=1}^S 
         \left(\Delta_{\rm gran,s} - \overline{\Delta_{\rm gran}}\right)^2},
\eeq
derived from the 11 individual snapshots. The quantity
$\sigma$ can serve to judge the significance of the abundance corrections
for each of the lines listed in the tables below.

%==============================================================================
\section{Results for the Sun}
\label{S5}
%==============================================================================
We have applied the above procedure to a representative sample of
lines, covering different elements, ionization stages, ionization
energies, excitation energies, line strengths and wavelengths.  In
this differential study, we have adopted the same representative van
der Waals broadening constant for all lines, $\log C_6=-30.3867$
(definition according to Mihalas (\cite{Mi78}), which differs 
from that by Uns\"old (\cite{U55}) by a factor $2\pi$), and
have neglected line broadening due to electron collisions, $C_4=0$.
The results are listed in Tables \ref{tab1}, \ref{tab2}, \ref{tab3},
and \ref{tab4}. Most of the lines are \emph{'fictitious' lines}
(`even' wavelength entries in the tables), allowing a systematic
study of the granulation effect as a function of the line
parameters.
The tables also contain a number of \emph{'real' lines} (table entries
with `uneven' wavelength) for Li, O, and Si, which have already been
used with detailed abundance studies by Wedemeyer (\cite{SW01}) and
Holweger (\cite{HH01}). In the latter work, the granulation
abundance corrections were obtained by interpolation in the list of
fictitious lines (Tables \ref{tab1},\ref{tab3}), which covers a
sufficient range of the relevant line parameters.

\subsection{Atoms with high ionization potential and ions}
\label{S51}
Atoms with high ionization potential like C\,{\sc i}, N\,{\sc i},
and O\,{\sc i}, as well as singly ionized elements are
representative of species in the main ionization stage.
Neutral nitrogen (ionization potential $\chi=14.5$~eV) can
be taken as the prototype of these species. Unlike C and 
O, N is not significantly affected by molecule formation 
or changes in the degree of ionization.

%=========================================================================
% Plot: Fig4
\begin{figure}
\resizebox{\hsize}{!}
{\mbox{\includegraphics[bb=00 56 538 400]{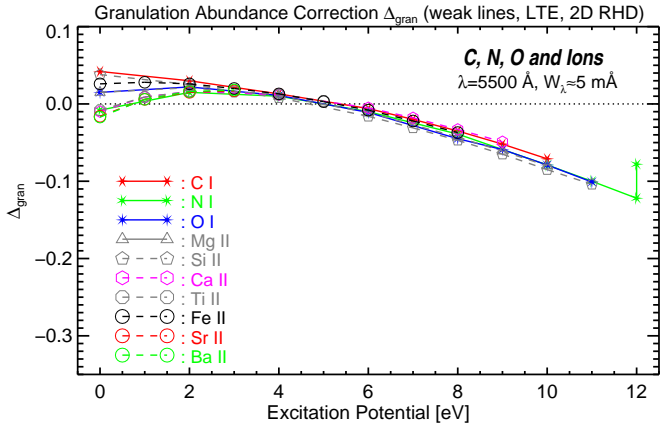}}}
\caption[]{%
   Logarithmic granulation abundance corrections $\Delta_{\rm gran}$ as a
   function of excitation energy $E_i$ for C\,{\sc i}, N\,{\sc i},
   O\,{\sc i}, and a number of singly ionized elements, representative of
   species in the main ionization stage. The plot is based on a subset of
   weak fictitious lines with $\lambda=5\,500$~\AA, $W_\lambda
   \approx 5$~m\AA, taken from Tables \ref{tab1}, \ref{tab2}, \ref{tab3},
   \ref{tab4}. The upper symbol at $E_i=12$~eV for N\,{\sc i} refers to 
   $\lambda\,10\,000$~\AA.
} % end of caption
\label{delta_ev_B1}
\end{figure}
%=========================================================================
% Plot: Fig5
\begin{figure}
\resizebox{\hsize}{!}
{\mbox{\includegraphics[bb=50 58 520 350]{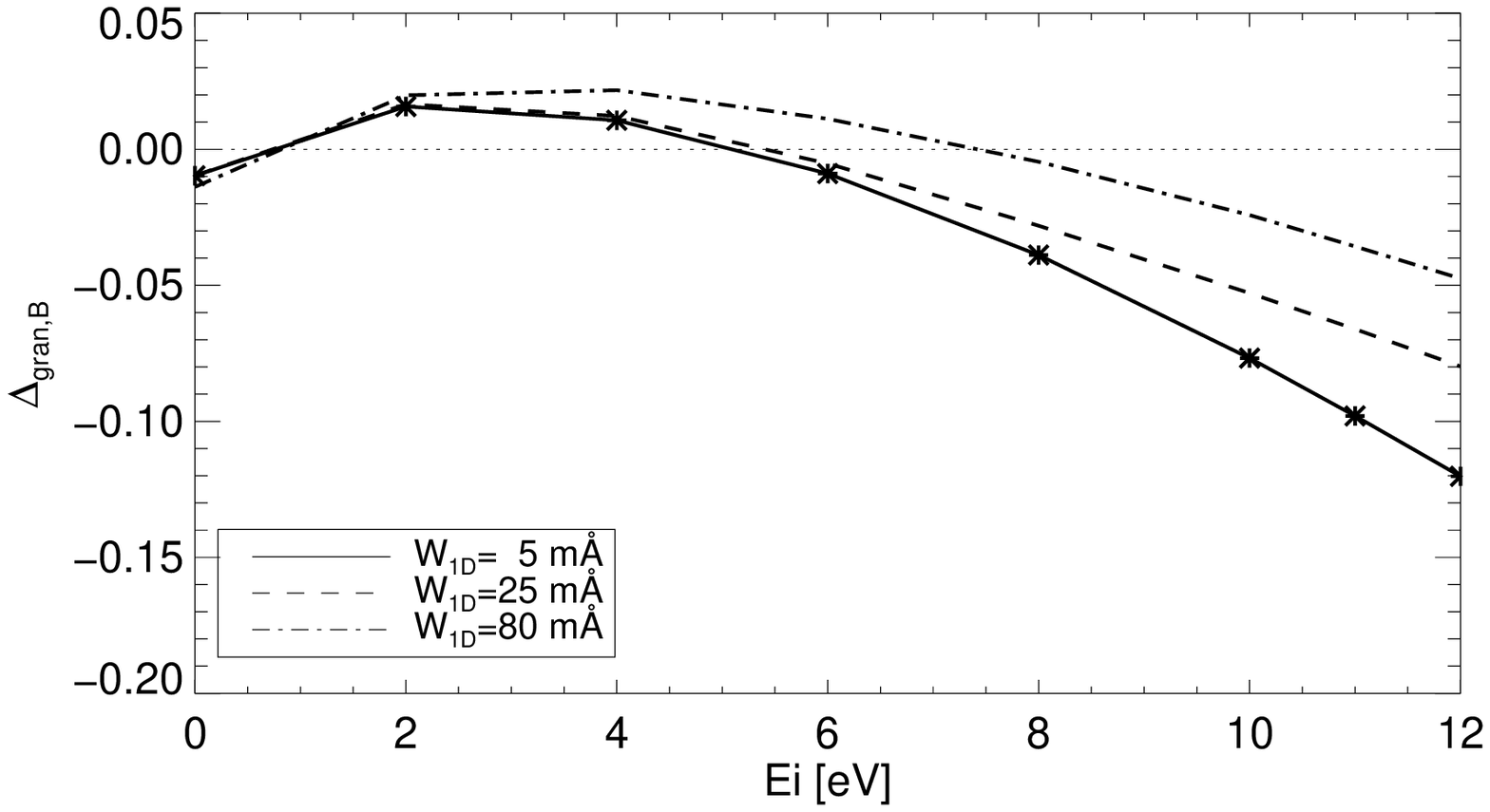}}}
\caption[]{%
\label{F05}
   Granulation abundance corrections, $\Delta_{\rm gran}$, for the
   first 8 fictitious nitrogen lines listed in Table~\ref{tab1}, 
   compared with test results obtained by increasing the line strength
   by factors of 5 and 15, respectively. 
} % end of caption
\end{figure}
%=========================================================================
% Plot: Fig6
\begin{figure}
\resizebox{\hsize}{!}
{\mbox{\includegraphics[bb=50 58 520 350]{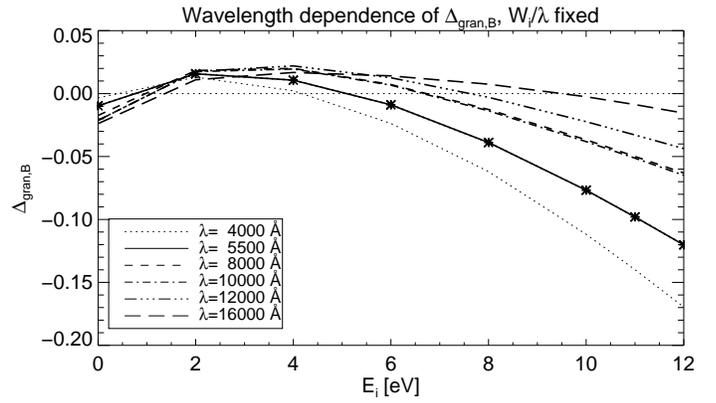}}}
\caption[]{%
\label{F06}
   Granulation abundance corrections, $\Delta_{\rm gran}$, for the
   first 8 fictitious nitrogen lines listed in Table~\ref{tab1}, 
   compared with test results obtained by moving the weak lines from
   $\lambda = 5\,500$~\AA\ to different spectral regions, centered
   at $\lambda=4\,000$, $8\,000$, $10\,000$, $12\,000$, and 
   $16\,000$~\AA, respectively, keeping the \emph{reduced} equivalent 
   width $W/\lambda$ fixed ($W_{\rm 1D} \approx \lambda/5\,500 \cdot 5$~m\AA).
} % end of caption
\end{figure}
%=========================================================================

The corrections $\Delta_{\rm gran}$ as a function of excitation
energy, $E_i$, are displayed in Fig.~\ref{delta_ev_B1} for our
tabulated sample of weak fictitious lines ($\lambda\,5500$~\AA, $W
\approx 5$~m\AA) of these species.  Obviously, the corrections for
high-excitation lines of the `majority species' are similar for all
elements, and can amount to -0.1 dex for weak lines with $E_i \ga
10$~eV.  This conclusion is independent of the assumed velocity field
which is irrelevant here: $\Delta_{\rm gran,A} \approx \Delta_{\rm
gran,B}$ for lines as weak as 5~m\AA\ (see tables).

\subsubsection{Dependence on line strength}
The dependence of the granulation abundance corrections on line
strength was investigated by test calculations for two additional
sets of stronger N\,{\sc i} lines. The two set of lines were obtained
from the set of N\,{\sc i} lines ($\lambda\, 5\,500$~\AA) listed in 
Table~\ref{tab1} by increasing their equivalent widths, $W_{\rm 1D}$,
by a factor of 5 and 15, respectively. Thus, the test lines in first
set of have $W_{\rm 1D} \approx 25$~m\AA, those in the second set
$W_{\rm 1D} \approx 80$~m\AA. According to the results shown in
Fig.~\ref{F05}, the granulation abundance corrections $\Delta_{\rm
gran}$ are systematically more positive for the stronger, otherwise
identical lines. For $E_i=12$~eV the difference in $\Delta_{\rm gran}$
between the lines with $W_{\rm 1D} \approx 80$~m\AA\ and $5$~m\AA\
amounts to +0.07~dex.

\subsubsection{Dependence on wavelength}
In order to get an idea about the dependence of the granulation 
abundance corrections on the wavelength of the spectral lines,
we also performed test calculations for five additional
sets of N\,{\sc i} lines, obtained
by shifting the weak N\,{\sc i} lines listed in Table~\ref{tab1} 
from $\lambda = 5\,500$~\AA\ to $\lambda = 4\,000$, $8\,000$, 
$10\,000$, $12\,000$, and $16\,000$~\AA, respectively, enforcing 
the same \emph{reduced} equivalent width, $W/\lambda$, i.e.
$W_{\rm 1D} \approx \lambda/5\,500 \cdot 5$~m\AA. 

The results shown in Fig.~\ref{F06} indicate that the granulation
abundance corrections $\Delta_{\rm gran}$ indeed depend on
wavelength, in the sense that lines in the blue part of the spectrum
show systematically more negative corrections than those in the red
part. This general trend is evident over the whole wavelength range,
apart from a slight reversal between $\lambda\, 8\,000$ and
$10\,000$~\AA\ which is related to the Paschen jump at $\lambda\,
8208$~\AA. For $E_i=12$~eV the difference in $\Delta_{\rm gran}$
between a weak line at $\lambda\, 4\,000$ and $10\,000$~\AA\ amounts
to -0.1~dex.

\subsection{Atoms with low ionization potential}
\label{S52}
Elements with low ionization potential are predominantly singly ionized
in the solar atmosphere. The particle density of the remaining neutral
atoms of such elements is particularly sensitive to temperature
variations. Molecules fall into the same category.
Neutral iron (ionization potential $\chi=7.9$~eV) can be taken 
as representative for these species. According to our models, the
fraction of neutral iron is nowhere exceeding 26\%; it shows a
maximum near $\log \tau_{5500}=-3$ and has dropped to less 
than 3.5\% at $\log \tau_{5500}=0$.

%=========================================================================
% Plot: Fig7
\begin{figure}
\resizebox{\hsize}{!}
{\mbox{\includegraphics[bb=00 56 538 400]{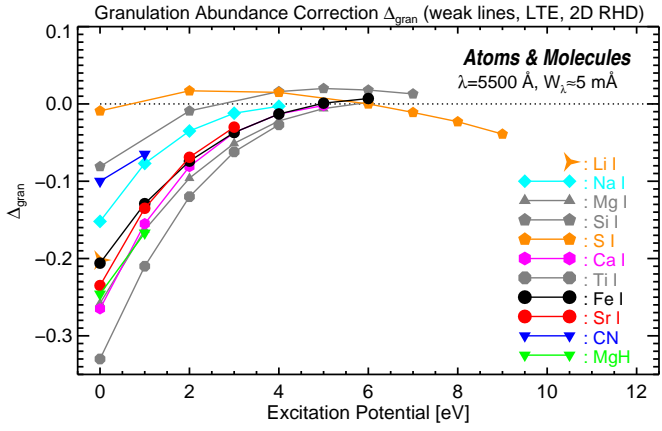}}}
\caption[]{%
   Logarithmic granulation abundance corrections $\Delta_{\rm gran}$ as a
   function of excitation energy $E_i$ for neutral atoms of predominantly
   ionized elements, and for molecules CN and MgH.
   The plot is based on a subset of weak fictitious lines with
   $\lambda=5\,500$~\AA, $W_\lambda \approx 5$~m\AA, taken from 
   Tables \ref{tab1}, \ref{tab2}, \ref{tab3}, \ref{tab4}.
} % end of caption
\label{delta_ev_B2}
\end{figure}
%=========================================================================
% Plot: Fig8
\begin{figure}
\resizebox{\hsize}{!}
{\mbox{\includegraphics[bb=50 58 520 350]{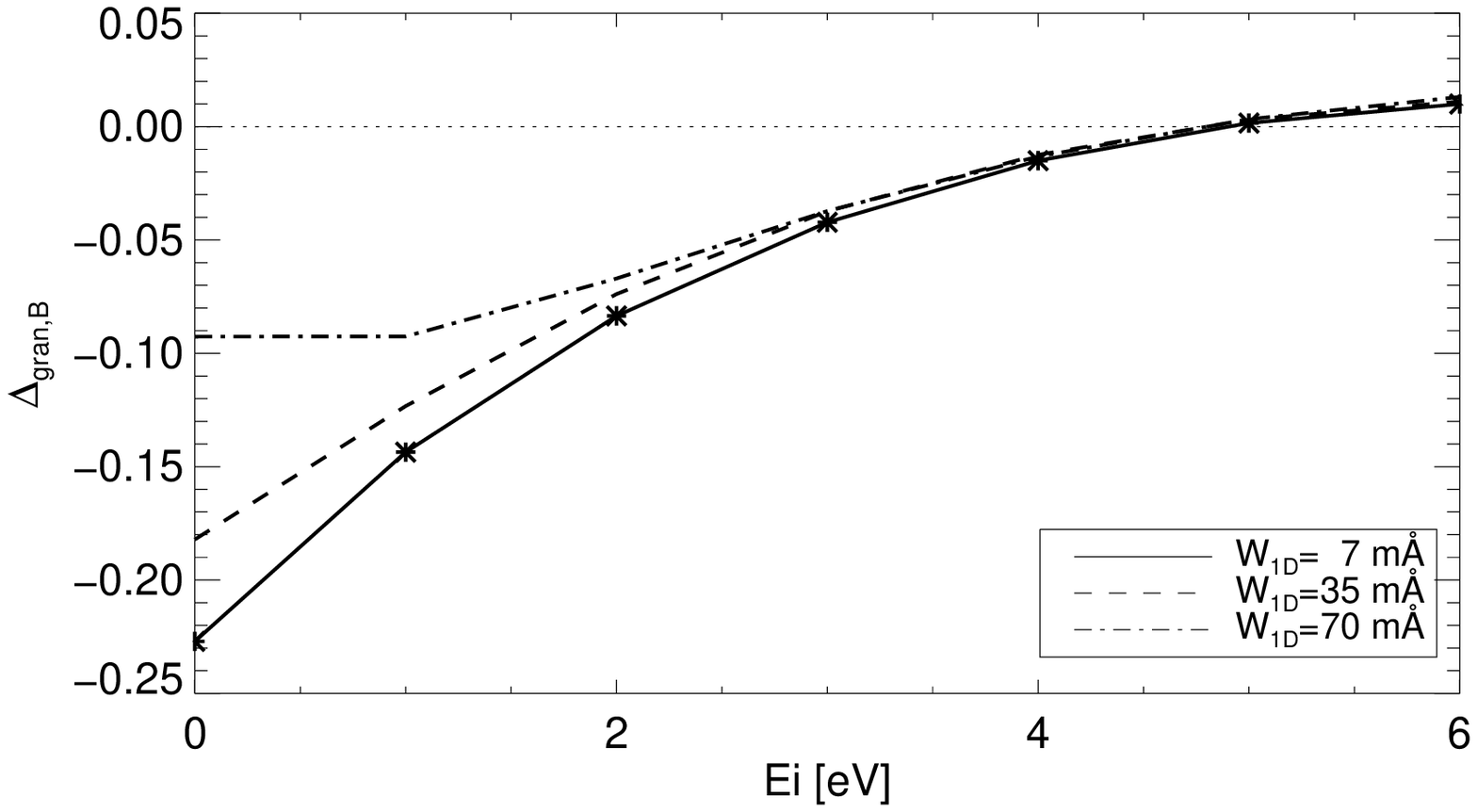}}}
\caption[]{%
\label{F08}
   Granulation abundance corrections, $\Delta_{\rm gran}$, for the
   7 fictitious neutral iron lines listed in Table~\ref{tab3}, 
   compared with test results obtained by increasing the line strength
   by factors of 5 and 10, respectively. 
} % end of caption
\end{figure}
%=========================================================================
% Plot: Fig9
\begin{figure}
\resizebox{\hsize}{!}
{\mbox{\includegraphics[bb=50 58 520 350]{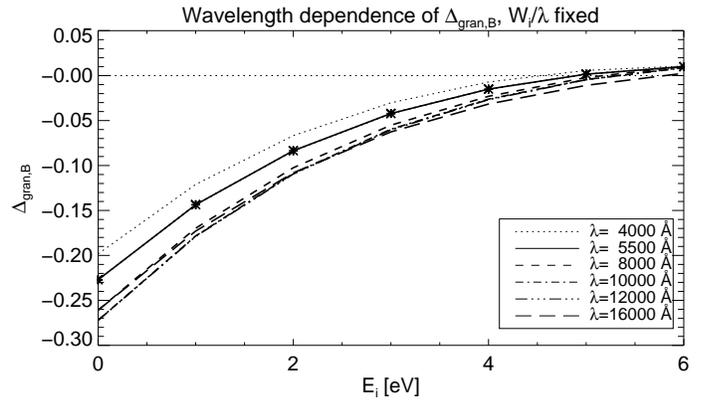}}}
\caption[]{%
\label{F09}
   Granulation abundance corrections, $\Delta_{\rm gran}$, for the
   7 fictitious neutral iron lines listed in Table~\ref{tab3}, 
   compared with test results obtained by moving the weak lines from
   $\lambda = 5\,500$~\AA\ to different spectral regions, centered
   at $\lambda=4\,000$, $8\,000$, $10\,000$, $12\,000$, and 
   $16\,000$~\AA, respectively, keeping the \emph{reduced} equivalent 
   width $W/\lambda$ fixed ($W_{\rm 1D} \approx \lambda/5\,500 \cdot 5$~m\AA).
} % end of caption
\end{figure}
%=========================================================================

The corrections $\Delta_{\rm gran}$ for a sample of weak fictitious
lines ($\lambda\,5500$~\AA, $W \approx 5$~m\AA) of these `minority
species' are displayed in Fig.~\ref{delta_ev_B2} as a function of
excitation energy, $E_i$.  We note that the largest granulation
effects are found for low-excitation lines of neutral atoms with an
ionization potential between 6 and 8~eV. The magnitude of the
correction depends on the element and is as large as -0.3 dex for
Ti\,{\sc i}, $E_i=0$~eV. Again, we point out that this result does not
depend on the assumed non-thermal velocity field.

It is interesting to note that Asplund (\cite{As00a}), using 3D
models, finds corrections in the opposite direction for the 'real' Si
I lines listed in Table~\ref{tab2}. We emphasize that this is not an
indication of fundamentally different ``granulation corrections'' in
2D and 3D.  The corrections given by Asplund with respect to the
Holweger-M\"uller (\cite{HM74}) model include not only the effect of
temperature inhomogeneities but also the influence of the different
mean stratifications.  Obviously, the difference in the mean structure
is the more important of the two opposing effects in this
case. Indeed, we find from our own models a \emph{total} correction of
-0.05 dex with respect to the Holweger-M\"uller model, compared to a
the tabulated "granulation correction" of +0.02 dex, explaining the
apparently inconsistent results.

\subsubsection{Dependence on line strength}
As for N\,{\sc i}, the dependence of the granulation abundance
corrections on line strength was investigated by test calculations for
two additional sets of stronger Fe\,{\sc i} lines, obtained from the set
of Fe\,{\sc i} lines listed in Table~\ref{tab3} by increasing their
equivalent widths, $W_{\rm 1D}$, by a factor of 5 and 10,
respectively. The test lines in first set of have then $W_{\rm 1D}
\approx 35$~m\AA, those in the second set $W_{\rm 1D} \approx
70$~m\AA. The results are shown in Fig.~\ref{F08}.  Qualitatively, the
granulation abundance corrections $\Delta_{\rm gran}$ for Fe\,{\sc i}
show the same dependence on line strength as already seen for N\,{\sc i}: 
they are systematically more positive for the stronger, otherwise
identical lines. 

The corrections for $E_i=0$ and $1$~eV, $W_{\rm 1D} \approx
70$~m\AA, shown in  Fig.~\ref{F08} are certainly too small (too 
positive), because a significant part of the line absorption
comes from layers located outside the model atmosphere on which
the line formation calculations are based ($\tau_{5500} < 10^{-5}$).
This situation results in a stronger underestimation of the equivalent
width for the inhomogeneous case. Hence, $\Delta_{\rm gran}$ should
actually be more negative, and we can expect abundance corrections
$\Delta_{\rm gran} < -0.1$~dex even for the stronger ground state lines.

\subsubsection{Dependence on wavelength}
We have performed test calculations for five additional sets of
Fe\,{\sc i} lines, obtained by shifting the standard Fe\,{\sc i} lines
listed in Table~\ref{tab3} from $\lambda = 5\,500$~\AA\ to $\lambda =
4\,000$, $8\,000$, $10\,000$, $12\,000$, and $16\,000$~\AA,
respectively, enforcing the same \emph{reduced} equivalent width,
$W/\lambda$, i.e.  $W_{\rm 1D} \approx \lambda/5\,500 \cdot 5$~m\AA.

The results shown in Fig.~\ref{F09} indicate that the wavelength
dependence of the granulation abundance corrections $\Delta_{\rm gran}$ 
for Fe\,{\sc i} has the opposite sign as in the case of N\,{\sc i}.
Lines in the blue part of the spectrum tend to show smaller (less negative)
corrections than those at longer wavelengths.
This general trend is clearly evident from $\lambda\, 4000$ to $8000$~\AA;
at longer wavelengths the variation of $\Delta_{\rm gran}$ with $\lambda$
is much reduced and no longer strictly monotonic (for an explanation
see Paper~II).

We have to point out that the results for
$\lambda$~16\,000~\AA\ are uncertain, because a significant part of
the line absorption comes from layers above $\tau_{16000} < 10^{-5}$,
so the line formation region is not entirely covered by our model
atmosphere. 

%==============================================================================
\section{Conclusions} \label{concl}
%==============================================================================

In an effort to estimate the \-- hitherto largely unknown \-- effects
of photospheric temperature fluctuations on spectroscopic abundance
determinations, we have carried out numerical simulations of the
problem of \emph{LTE line formation in the inhomogeneous solar
photosphere} based on detailed 2-dimensional radiation hydrodynamics
models of the convective surface layers of the Sun.

For a variety of spectral lines of different elements, we have computed
synthetic line profiles from the inhomogeneous 2D hydrodynamical atmosphere 
and from the corresponding 1D plane-parallel model, respectively.
By means of a strictly differential 1D/2D comparison of the emergent
equivalent widths, we have derived so called ``granulation abundance
corrections'' for the individual lines, which have to be applied to
standard abundance determinations based on homogeneous 1D model atmospheres
in order to correct for the influence of the photospheric temperature
fluctuations. The `classical' problem of finding the most appropriate
mean vertical temperature stratification is \emph{not} addressed here.

Using the concept of ``fictitious'' spectral lines, we were able to
investigate systematically the dependence of the ``granulation
abundance corrections'' on the basic line parameters like excitation
potential, line strength, wavelength, element and ionization
stage. For many lines of practical relevance, it should be possible to
estimate the magnitude of the abundance correction by interpolation in
the graphs and tables provided in this paper. This
approach may often be an acceptable alternative to a detailed fitting
of individual line profiles based on hydrodynamical simulations.

In general, we find a \emph{line strengthening} in the presence of
temperature inhomogeneities, implying mostly \emph{negative}
``granulation abundance corrections'', i.e. standard analysis based on
plane-parallel atmospheres tends to \emph{overestimate} abundances.
The physical reason for the line strengthening is primarily the
non-linear temperature dependence of the line opacity due to thermal
ionization and excitation, as demonstrated in Paper~II.

One remarkable result is that all lines of our sample with an
excitation potential around $E_i \approx 5$~eV are practically
\emph{insensitive} to granulation effects, regardless
of element and ionization stage.
Moderate granulation corrections ($\Delta_{\rm gran} \approx
-0.1$~dex) are found for weak, high-excitation lines ($E_i \ga 10$~eV)
of ions and atoms with high ionization potential like N\,{\sc i}.  The
largest corrections are found for ground state lines ($E_i=0$~eV) of
neutral atoms with an ionization potential between 6 and 8~eV like
Mg\,{\sc i}, Ca\,{\sc i}, Ti\,{\sc i}, Fe\,{\sc i}, amounting to
$\Delta_{\rm gran} \approx -0.3$~dex in the case of Ti\,{\sc i}.

For given excitation potential, the granulation corrections are
systematically more positive for stronger lines. The wavelength
dependence of $\Delta_{\rm gran}$, however, depends on the type of
line (`majority' or `minority' species).

We recall that the corrections $\Delta_{\rm gran}$ account
only for the influence of the temperature and pressure fluctuations
relative to a \emph{given mean structure}. We have
suppressed possible effects related to 1D/2D differences in the
details of the non-thermal velocity field. Such additional effects,
which apply to partly saturated lines only, should be small, however,
since they are largely eliminated by the proper choice of the
microturbulence parameter $\xi_{\rm mic}$.

We point out that the magnitude of the ``granulation abundance
corrections'' derived in this work should be considered as upper
limits ($|\Delta_{\rm gran}|(2D,LTE) > |\Delta_{\rm gran}|(3D,NLTE)$),
because (i) possible NLTE-effects, which were ignored in this study,
tend to reduce the fluctuations of the line
opacity (e.g.\ Kiselman \cite{Ki97}, Cayrel \& Steffen \cite{CS00}, Asplund
\cite{As00b}), and (ii) the amplitude of the temperature fluctuations is
systematically overestimated in 2D relative to 3D convection models 
(see Fig.~\ref{Trms}). The large negative corrections found for the
low-excitation lines of atoms with low ionization potential are the
result of the strongly increasing amplitude of the temperature
fluctuations towards the upper photosphere as predicted by our present
2D hydrodynamical simulations. In a more realistic 3D atmosphere, this
feature will therefore be less pronounced.

Even though the results are still somewhat uncertain, the present study
should be helpful in providing upper bounds for possible errors of
spectroscopic abundance analyses, and for selecting spectral lines which
are as insensitive as possible to the effects of photospheric temperature
inhomogeneities.

For a physical explanation of the numerical results obtained in this
work, a detailed investigation of the process of LTE line formation in an
inhomogeneous stellar atmosphere based on the analysis of the transfer
equation and the evaluation of line depression contribution functions
is presented in the second paper of this series (Paper~II).  With the
help of this formalism, we can actually tell how the different
atmospheric layers contribute to the granulation correction for any
particular spectral line.

In the near future, we plan to extend the computation of abundance
corrections to full 3D simulations for the Sun and other types of
stars where convection extends into higher photospheric layers and the
spectroscopic granulation effects may well be even larger.

\begin{acknowledgements}
The 2D numerical convection simulations and line formation
calculations were carried out on the CRAY T94 and CRAY SV1,
respectively, at the \emph{Rechenzentrum der Universit\"at Kiel}.
We thank the referee, Paul Barklem, for constructive criticism which
helped to significantly improve the contents and presentation of 
this work.
\end{acknowledgements}

\clearpage
%==============================================================================

%==============================================================================
\section{Granulation Abundance Correction Tables}
%==============================================================================
\begin{table*}[htb]
\begin{center}
\boldmath
\begin{tabular}{|r|r|r|r|r r|r r|r|} \hline
{\bfseries Ion} & $\lambda$ [\AA] & $E_{\rm i}$ [eV] & $W_{\rm 1D}$ [m\AA] & 
\multicolumn{2}{c|}{$W_{\rm 2D}$ [m\AA]} &
\multicolumn{2}{c|}{$\Delta_{\rm gran}$} & $\sigma$ \\
\hline\hline

Li\,{\sc i} &  6707.8 &   0.0 &   7.571 & 11.835 & (11.828) & -0.202 & (-0.201) & 0.088 \\
\hline\hline				                                   
 C\,{\sc i} &  5500.0 &   0.0 &   5.036 &  4.580 &  (4.575) & +0.042 & (+0.043) & 0.010 \\
 C\,{\sc i} &  5500.0 &   2.0 &   4.981 &  4.658 &  (4.652) & +0.030 & (+0.030) & 0.009 \\
 C\,{\sc i} &  5500.0 &   4.0 &   5.016 &  4.859 &  (4.851) & +0.013 & (+0.014) & 0.007 \\
 C\,{\sc i} &  5500.0 &   6.0 &   5.116 &  5.188 &  (5.174) & -0.006 & (-0.005) & 0.006 \\
 C\,{\sc i} &  5500.0 &   7.0 &   5.115 &  5.334 &  (5.316) & -0.020 & (-0.018) & 0.007 \\
 C\,{\sc i} &  5500.0 &   8.0 &   5.184 &  5.577 &  (5.554) & -0.035 & (-0.033) & 0.008 \\
 C\,{\sc i} &  5500.0 &   9.0 &   5.282 &  5.878 &  (5.848) & -0.052 & (-0.050) & 0.010 \\
 C\,{\sc i} &  5500.0 &  10.0 &   5.377 &  6.197 &  (6.157) & -0.071 & (-0.068) & 0.012 \\
\hline					                                   
 C\,{\sc i} &  9800.0 &   7.8 &  84.806 & 82.877 & (81.104) & +0.022 & (+0.043) & 0.016 \\
 C\,{\sc i} & 12700.0 &   8.7 &  80.479 & 78.720 & (77.264) & +0.019 & (+0.035) & 0.007 \\
\hline\hline				                                   
					                                   
 N\,{\sc i} &  5500.0 &   0.0 &   5.521 &  5.625 &  (5.619) & -0.008 & (-0.008) & 0.006 \\
 N\,{\sc i} &  5500.0 &   2.0 &   5.231 &  5.058 &  (5.051) & +0.015 & (+0.016) & 0.004 \\
 N\,{\sc i} &  5500.0 &   4.0 &   5.121 &  5.013 &  (5.002) & +0.010 & (+0.011) & 0.006 \\
 N\,{\sc i} &  5500.0 &   6.0 &   5.153 &  5.266 &  (5.248) & -0.010 & (-0.008) & 0.006 \\
 N\,{\sc i} &  5500.0 &   8.0 &   5.288 &  5.754 &  (5.723) & -0.039 & (-0.037) & 0.009 \\
 N\,{\sc i} &  5500.0 &  10.0 &   5.471 &  6.390 &  (6.335) & -0.079 & (-0.075) & 0.013 \\
 N\,{\sc i} &  5500.0 &  11.0 &   5.498 &  6.650 &  (6.582) & -0.100 & (-0.095) & 0.016 \\
 N\,{\sc i} &  5500.0 &  12.0 &   5.579 &  6.970 &  (6.885) & -0.122 & (-0.115) & 0.019 \\
\hline					                                   
 N\,{\sc i} &  9600.0 &  11.1 &   3.783 &  4.336 &  (4.304) & -0.067 & (-0.063) & 0.014 \\
 N\,{\sc i} & 10000.0 &  12.0 &   5.016 &  5.826 &  (5.767) & -0.078 & (-0.073) & 0.015 \\
\hline\hline				                                   
					                                   
 O\,{\sc i} &  5500.0 &   0.0 &   5.253 &  5.076 &  (5.069) & +0.015 & (+0.016) & 0.003 \\
 O\,{\sc i} &  5500.0 &   2.0 &   5.088 &  4.847 &  (4.838) & +0.022 & (+0.023) & 0.007 \\
 O\,{\sc i} &  5500.0 &   4.0 &   5.053 &  4.937 &  (4.925) & +0.011 & (+0.012) & 0.007 \\
 O\,{\sc i} &  5500.0 &   6.0 &   5.137 &  5.262 &  (5.240) & -0.011 & (-0.009) & 0.007 \\
 O\,{\sc i} &  5500.0 &   8.0 &   5.281 &  5.762 &  (5.725) & -0.045 & (-0.042) & 0.009 \\
 O\,{\sc i} &  5500.0 &   9.0 &   5.363 &  6.059 &  (6.010) & -0.059 & (-0.055) & 0.011 \\
 O\,{\sc i} &  5500.0 &  10.0 &   5.462 &  6.388 &  (6.322) & -0.079 & (-0.073) & 0.014 \\
 O\,{\sc i} &  5500.0 &  11.0 &   5.563 &  6.725 &  (6.642) & -0.101 & (-0.095) & 0.016 \\
\hline					                                   
 O\,{\sc i} &  5500.0 &  10.0 &   5.462 &  6.388 &  (6.322) & -0.079 & (-0.073) & 0.014 \\
 O\,{\sc i} &  5500.0 &  10.0 &  30.801 & 32.812 & (31.956) & -0.052 & (-0.030) & 0.012 \\
 O\,{\sc i} &  5500.0 &  10.0 &  60.556 & 62.594 & (60.707) & -0.035 & (-0.003) & 0.012 \\
 O\,{\sc i} &  5500.0 &  10.0 &  88.896 & 90.792 & (88.154) & -0.025 & (+0.010) & 0.014 \\
\hline					                                   
 O\,{\sc i} &  5500.0 &  10.0 &  60.556 & 62.594 & (60.707) & -0.035 & (-0.003) & 0.012 \\
 O\,{\sc i} &  7500.0 &  10.0 &  56.869 & 57.203 & (55.610) & -0.006 & (+0.022) & 0.013 \\
 O\,{\sc i} &  9500.0 &  10.0 &  56.959 & 57.295 & (55.840) & -0.005 & (+0.018) & 0.012 \\
 O\,{\sc i} & 11500.0 &  10.0 &  58.720 & 58.650 & (57.281) & +0.001 & (+0.021) & 0.008 \\
\hline					                                   
% O\,{\sc i} &  7774.2 &  9.15 &  66.214 & 65.619 & (63.773) & +0.009 & (+0.038) & 0.014 \\
 O\,{\sc i} &  8700.0 &  8.80 &  33.827 & 34.123 & (33.344) & -0.006 & (+0.011) & 0.010 \\
\hline					                                   
 O\,{\sc i} &  6158.2 & 10.74 &   5.408 &  6.375 &  (6.336) & -0.081 & (-0.078) & 0.014 \\
 O\,{\sc i} &  6300.3 &  0.00 &   4.479 &  4.329 &  (4.324) & +0.015 & (+0.016) & 0.003 \\
 O\,{\sc i} &  7771.9 &  9.15 &  82.207 & 81.184 & (79.130) & +0.013 & (+0.039) & 0.015 \\
 O\,{\sc i} &  7774.2 &  9.15 &  67.155 & 66.689 & (64.988) & +0.007 & (+0.031) & 0.013 \\
 O\,{\sc i} &  7775.4 &  9.15 &  51.250 & 51.273 & (50.002) & -0.000 & (+0.021) & 0.012 \\
 O\,{\sc i} &  9265.9 & 10.74 &  34.068 & 35.492 & (35.062) & -0.026 & (-0.018) & 0.009 \\
 O\,{\sc i} & 11302.4 & 10.74 &  14.074 & 14.947 & (14.807) & -0.033 & (-0.028) & 0.008 \\
 O\,{\sc i} & 13164.9 & 10.99 &  17.303 & 17.894 & (17.705) & -0.017 & (-0.013) & 0.006 \\
\hline\hline
\end{tabular}
\unboldmath
\caption[]
{
Granulation abundance corrections for Li\,{\sc i}, C\,{\sc i},
N\,{\sc i}, and O\,{\sc i}, derived from a sample of representative
`fictitious' lines (even wavelengths) and some selected `real' lines
commonly used in the analysis of late type stars. $E_{\rm i}$ is the
excitation potential of the line's lower energy level, $W_{\rm 2D}$ is
the mean equivalent width computed from 11 selected 2D snapshots,
replacing the hydrodynamical velocity field by a depth-independent
mircroturbulence of $\xi_{\rm mic}=1$~km/s; $W_{\rm 1D}$ is the mean
equivalent width computed from the corresponding 1D models ($\xi_{\rm
mic}=1$~km/s). $\Delta_{\rm gran}$ is the resulting granulation
abundance correction derived from the ratio $W_{\rm 2D}/W_{\rm 1D}$,
including curve-of-growth effects.  For reference, we give in
parentheses the results obtained when the 2D line profiles are
computed with the hydrodynamical velocity field of the respective
snapshot; these corrections, however, are not recommended (see text
for details). $\sigma$ is the standard deviation of the abundance
corrections derived from the 11 individual snapshots, measuring the 
statistical uncertainty of the tabulated $\Delta_{\rm gran}$.
}
\label{tab1}
\end{center}
\end{table*}

\clearpage

\begin{table*}[htb]
\begin{center}
\boldmath
\begin{tabular}{|r|r|r|r|r r|r r|r|} \hline
{\bfseries Ion} & $\lambda$ [\AA] & $E_{\rm i}$ [eV] & $W_{\rm 1D}$ [m\AA] & 
\multicolumn{2}{c|}{$W_{\rm 2D}$ [m\AA]} &
\multicolumn{2}{c|}{$\Delta_{\rm gran}$} & $\sigma$ \\
\hline\hline

Na\,{\sc i} &  5500.0 &   0.0 &   7.357 & 10.186 & (10.147) & -0.152 & (-0.150) & 0.063 \\
Na\,{\sc i} &  5500.0 &   1.0 &   6.874 &  8.131 &  (8.104) & -0.077 & (-0.076) & 0.030 \\
Na\,{\sc i} &  5500.0 &   2.0 &   6.482 &  6.993 &  (6.969) & -0.035 & (-0.033) & 0.014 \\
Na\,{\sc i} &  5500.0 &   3.0 &   6.202 &  6.370 &  (6.346) & -0.012 & (-0.011) & 0.006 \\
Na\,{\sc i} &  5500.0 &   4.0 &   5.992 &  6.030 &  (6.003) & -0.003 & (-0.001) & 0.003 \\
\hline\hline			                 	                  
				                 	                  
Mg\,{\sc i} &  5500.0 &   0.0 &   8.783 & 15.107 & (14.974) & -0.258 & (-0.254) & 0.087 \\
Mg\,{\sc i} &  5500.0 &   1.0 &   8.034 & 11.392 & (11.327) & -0.163 & (-0.161) & 0.054 \\
Mg\,{\sc i} &  5500.0 &   2.0 &   7.425 &  9.142 &  (9.101) & -0.096 & (-0.094) & 0.032 \\
Mg\,{\sc i} &  5500.0 &   3.0 &   6.937 &  7.754 &  (7.723) & -0.051 & (-0.049) & 0.019 \\
Mg\,{\sc i} &  5500.0 &   4.0 &   6.545 &  6.870 &  (6.843) & -0.022 & (-0.020) & 0.011 \\
Mg\,{\sc i} &  5500.0 &   5.0 &   6.239 &  6.312 &  (6.287) & -0.006 & (-0.004) & 0.006 \\
Mg\,{\sc i} &  5500.0 &   6.0 &   6.011 &  5.982 &  (5.955) & +0.002 & (+0.004) & 0.003 \\
\hline\hline			                 	                  
				                 	                  
Mg\,{\sc ii} &  5500.0 &  0.0 &   5.394 &  5.221 &  (5.207) & +0.015 & (+0.016) & 0.005 \\
Mg\,{\sc ii} &  5500.0 &  2.0 &   5.153 &  4.920 &  (4.905) & +0.021 & (+0.022) & 0.006 \\
Mg\,{\sc ii} &  5500.0 &  4.0 &   5.084 &  4.970 &  (4.949) & +0.010 & (+0.012) & 0.006 \\
Mg\,{\sc ii} &  5500.0 &  6.0 &   5.138 &  5.264 &  (5.229) & -0.011 & (-0.008) & 0.007 \\
Mg\,{\sc ii} &  5500.0 &  8.0 &   5.269 &  5.743 &  (5.681) & -0.043 & (-0.037) & 0.009 \\
Mg\,{\sc ii} &  5500.0 &  9.0 &   5.358 &  6.063 &  (5.954) & -0.061 & (-0.054) & 0.011 \\
Mg\,{\sc ii} &  5500.0 & 10.0 &   5.438 &  6.328 &  (6.223) & -0.080 & (-0.071) & 0.014 \\
\hline				                 	                  
Mg\,{\sc ii} &  8800.0 &  9.4 &  42.424 & 42.579 & (41.078) & -0.003 & (+0.027) & 0.011 \\
\hline\hline			                 	                  
				                 	                  
Si\,{\sc i} &  5500.0 &   0.0 &   7.398 &  8.814 &  (8.768) & -0.081 & (-0.079) & 0.033 \\
Si\,{\sc i} &  5500.0 &   2.0 &   6.538 &  6.664 &  (6.635) & -0.009 & (-0.007) & 0.012 \\
Si\,{\sc i} &  5500.0 &   4.0 &   6.015 &  5.801 &  (5.776) & +0.016 & (+0.018) & 0.006 \\
Si\,{\sc i} &  5500.0 &   5.0 &   5.840 &  5.595 &  (5.569) & +0.020 & (+0.022) & 0.005 \\
Si\,{\sc i} &  5500.0 &   6.0 &   5.709 &  5.490 &  (5.461) & +0.018 & (+0.021) & 0.005 \\
Si\,{\sc i} &  5500.0 &   7.0 &   5.622 &  5.473 &  (5.439) & +0.013 & (+0.016) & 0.005 \\
\hline							                   
							                   
Si\,{\sc i} &  5645.6 & 4.9 &  38.063 &  36.710 &  (36.131) & +0.022 & (+0.031) & 0.006 \\
Si\,{\sc i} &  5708.4 & 5.0 &  90.285 &  88.150 &  (85.562) & +0.023 & (+0.051) & 0.006 \\
Si\,{\sc i} &  5772.1 & 5.1 &  56.938 &  55.137 &  (54.016) & +0.023 & (+0.037) & 0.006 \\
Si\,{\sc i} &  5793.1 & 4.9 &  43.616 &  42.097 &  (41.335) & +0.023 & (+0.034) & 0.006 \\
Si\,{\sc i} &  5948.5 & 5.1 & 104.552 & 102.096 &  (99.151) & +0.024 & (+0.052) & 0.006 \\
Si\,{\sc i} &  6976.5 & 6.0 &  46.790 &  44.704 &  (44.433) & +0.025 & (+0.028) & 0.006 \\
Si\,{\sc i} &  7932.3 & 6.0 & 129.359 & 125.041 & (122.930) & +0.029 & (+0.043) & 0.006 \\
Si\,{\sc i} &  7970.3 & 6.0 &  27.701 &  26.280 &  (26.125) & +0.027 & (+0.030) & 0.006 \\
\hline\hline						                   
							                   
Si\,{\sc ii} & 5500.0 &   0.0 &   4.824 &  4.430 &  (4.418) & +0.038 & (+0.040) & 0.011 \\
Si\,{\sc ii} & 5500.0 &   2.0 &   4.830 &  4.570 &  (4.553) & +0.025 & (+0.027) & 0.010 \\
Si\,{\sc ii} & 5500.0 &   4.0 &   4.921 &  4.842 &  (4.816) & +0.008 & (+0.010) & 0.008 \\
Si\,{\sc ii} & 5500.0 &   6.0 &   5.057 &  5.234 &  (5.190) & -0.016 & (-0.012) & 0.007 \\
Si\,{\sc ii} & 5500.0 &   7.0 &   5.136 &  5.474 &  (5.415) & -0.031 & (-0.026) & 0.008 \\
Si\,{\sc ii} & 5500.0 &   8.0 &   5.232 &  5.753 &  (5.676) & -0.047 & (-0.041) & 0.010 \\
Si\,{\sc ii} & 5500.0 &   9.0 &   5.325 &  6.044 &  (5.943) & -0.065 & (-0.057) & 0.012 \\
Si\,{\sc ii} & 5500.0 &  10.0 &   5.415 &  6.339 &  (6.211) & -0.085 & (-0.074) & 0.014 \\
Si\,{\sc ii} & 5500.0 &  11.0 &   5.506 &  6.633 &  (6.476) & -0.104 & (-0.091) & 0.017 \\
\hline				                 	                  
Si\,{\sc ii} & 6347.1 &  8.12 &  47.386 & 48.453 & (46.572) & -0.021 & (+0.016) & 0.013 \\
Si\,{\sc ii} & 6371.4 &  8.12 &  33.817 & 34.876 & (33.622) & -0.025 & (+0.005) & 0.011 \\
\hline\hline
\end{tabular}
\unboldmath
\caption[]
{
Same as Table~\ref{tab1}, but for a sample of Na\,{\sc i}, 
Mg\,{\sc i}, Mg\,{\sc ii}, Si\,{\sc i}, and Si\,{\sc ii} lines. 
}
\label{tab2}
\end{center}
\end{table*}

\begin{table*}[htb]
\begin{center}
\boldmath
\begin{tabular}{|r|r|r|r|r r|r r|r|} \hline
{\bfseries Ion} & $\lambda$ [\AA] & $E_{\rm i}$ [eV] & $W_{\rm 1D}$ [m\AA] & 
\multicolumn{2}{c|}{$W_{\rm 2D}$ [m\AA]} &
\multicolumn{2}{c|}{$\Delta_{\rm gran}$} & $\sigma$ \\
\hline\hline

 S\,{\sc i} &  5500.0 &   0.0 &   5.547 &  5.662 &  (5.642) & -0.009 & (-0.008) & 0.006 \\
 S\,{\sc i} &  5500.0 &   2.0 &   5.245 &  5.052 &  (5.031) & +0.017 & (+0.019) & 0.005 \\
 S\,{\sc i} &  5500.0 &   4.0 &   5.133 &  4.967 &  (4.940) & +0.015 & (+0.018) & 0.006 \\
 S\,{\sc i} &  5500.0 &   5.0 &   5.148 &  5.145 &  (5.103) & +0.000 & (+0.004) & 0.006 \\
 S\,{\sc i} &  5500.0 &   7.0 &   5.191 &  5.308 &  (5.254) & -0.011 & (-0.006) & 0.006 \\
 S\,{\sc i} &  5500.0 &   8.0 &   5.244 &  5.504 &  (5.436) & -0.023 & (-0.017) & 0.007 \\
 S\,{\sc i} &  5500.0 &   9.0 &   5.308 &  5.731 &  (5.643) & -0.039 & (-0.031) & 0.008 \\
\hline\hline			                 	                      
				                 	                      
Ca\,{\sc i} &  5500.0 &   0.0 &   8.006 & 13.864 & (13.649) & -0.265 & (-0.257) & 0.111 \\
Ca\,{\sc i} &  5500.0 &   1.0 &   7.401 & 10.271 & (10.179) & -0.155 & (-0.151) & 0.063 \\
Ca\,{\sc i} &  5500.0 &   2.0 &   6.917 &  8.227 &  (8.167) & -0.081 & (-0.078) & 0.031 \\
Ca\,{\sc i} &  5500.0 &   3.0 &   6.526 &  7.071 &  (7.021) & -0.037 & (-0.034) & 0.015 \\
Ca\,{\sc i} &  5500.0 &   4.0 &   6.221 &  6.400 &  (6.353) & -0.013 & (-0.010) & 0.006 \\
Ca\,{\sc i} &  5500.0 &   5.0 &   6.002 &  6.028 &  (5.978) & -0.002 & (+0.002) & 0.003 \\
\hline\hline			                 	                      
				                 	                      
Ca\,{\sc ii} & 5500.0 &   0.0 &   5.529 &  5.650 &  (5.624) & -0.010 & (-0.008) & 0.006 \\
Ca\,{\sc ii} & 5500.0 &   1.0 &   5.362 &  5.263 &  (5.237) & +0.009 & (+0.011) & 0.002 \\
Ca\,{\sc ii} & 5500.0 &   2.0 &   5.236 &  5.059 &  (5.032) & +0.016 & (+0.018) & 0.004 \\
Ca\,{\sc ii} & 5500.0 &   4.0 &   5.106 &  4.974 &  (4.938) & +0.012 & (+0.015) & 0.006 \\
Ca\,{\sc ii} & 5500.0 &   6.0 &   5.132 &  5.191 &  (5.134) & -0.005 & (-0.000) & 0.006 \\
Ca\,{\sc ii} & 5500.0 &   7.0 &   5.177 &  5.373 &  (5.300) & -0.018 & (-0.011) & 0.007 \\
Ca\,{\sc ii} & 5500.0 &   8.0 &   5.231 &  5.588 &  (5.492) & -0.033 & (-0.024) & 0.008 \\
Ca\,{\sc ii} & 5500.0 &   9.0 &   5.313 &  5.847 &  (5.723) & -0.049 & (-0.038) & 0.010 \\
\hline\hline			                 	                      
				                 	                      
Ti\,{\sc i} &  5500.0 &   0.0 &   8.503 & 16.635 & (16.169) & -0.330 & (-0.315) & 0.128 \\
Ti\,{\sc i} &  5500.0 &   1.0 &   7.816 & 12.113 & (11.929) & -0.210 & (-0.203) & 0.082 \\
Ti\,{\sc i} &  5500.0 &   2.0 &   7.251 &  9.346 &  (9.249) & -0.120 & (-0.115) & 0.045 \\
Ti\,{\sc i} &  5500.0 &   3.0 &   6.793 &  7.750 &  (7.680) & -0.062 & (-0.058) & 0.023 \\
Ti\,{\sc i} &  5500.0 &   4.0 &   6.432 &  6.817 &  (6.757) & -0.027 & (-0.023) & 0.011 \\
\hline\hline			                 	                      
				                 	                      
Ti\,{\sc ii} & 5500.0 &   0.0 &   5.526 &  5.619 &  (5.586) & -0.008 & (-0.005) & 0.005 \\
Ti\,{\sc ii} & 5500.0 &   1.0 &   5.353 &  5.244 &  (5.212) & +0.009 & (+0.012) & 0.002 \\
Ti\,{\sc ii} & 5500.0 &   2.0 &   5.224 &  5.044 &  (5.010) & +0.016 & (+0.019) & 0.005 \\
Ti\,{\sc ii} & 5500.0 &   3.0 &   5.146 &  4.969 &  (4.932) & +0.016 & (+0.020) & 0.006 \\
Ti\,{\sc ii} & 5500.0 &   4.0 &   5.114 &  4.988 &  (4.944) & +0.012 & (+0.016) & 0.006 \\
Ti\,{\sc ii} & 5500.0 &   5.0 &   5.108 &  5.067 &  (5.012) & +0.004 & (+0.009) & 0.006 \\
\hline\hline						                       
							                       
Fe\,{\sc i} &  5500.0 &   0.0 &   8.628 & 13.189 & (12.910) & -0.206 & (-0.195) & 0.070 \\
Fe\,{\sc i} &  5500.0 &   1.0 &   7.933 & 10.386 & (10.225) & -0.129 & (-0.121) & 0.044 \\
Fe\,{\sc i} &  5500.0 &   2.0 &   7.358 &  8.608 &  (8.498) & -0.074 & (-0.068) & 0.027 \\
Fe\,{\sc i} &  5500.0 &   3.0 &   6.899 &  7.468 &  (7.384) & -0.037 & (-0.032) & 0.017 \\
Fe\,{\sc i} &  5500.0 &   4.0 &   6.517 &  6.703 &  (6.632) & -0.013 & (-0.008) & 0.010 \\
Fe\,{\sc i} &  5500.0 &   5.0 &   6.219 &  6.205 &  (6.139) & +0.001 & (+0.006) & 0.006 \\
Fe\,{\sc i} &  5500.0 &   6.0 &   5.995 &  5.899 &  (5.832) & +0.007 & (+0.013) & 0.004 \\
\hline\hline			                 	                      
				                 	                      
Fe\,{\sc ii} & 5500.0 &   0.0 &   5.265 &  4.975 &  (4.941) & +0.026 & (+0.029) & 0.006 \\
Fe\,{\sc ii} & 5500.0 &   1.0 &   5.152 &  4.849 &  (4.813) & +0.028 & (+0.031) & 0.007 \\
Fe\,{\sc ii} & 5500.0 &   2.0 &   5.076 &  4.800 &  (4.761) & +0.026 & (+0.029) & 0.007 \\
Fe\,{\sc ii} & 5500.0 &   3.0 &   5.040 &  4.821 &  (4.777) & +0.020 & (+0.025) & 0.007 \\
Fe\,{\sc ii} & 5500.0 &   4.0 &   5.031 &  4.891 &  (4.838) & +0.013 & (+0.018) & 0.007 \\
Fe\,{\sc ii} & 5500.0 &   5.0 &   5.058 &  5.020 &  (4.953) & +0.003 & (+0.010) & 0.007 \\
Fe\,{\sc ii} & 5500.0 &   6.0 &   5.095 &  5.181 &  (5.096) & -0.008 & (-0.000) & 0.007 \\
Fe\,{\sc ii} & 5500.0 &   7.0 &   5.155 &  5.386 &  (5.276) & -0.022 & (-0.011) & 0.007 \\
Fe\,{\sc ii} & 5500.0 &   8.0 &   5.224 &  5.618 &  (5.475) & -0.037 & (-0.024) & 0.009 \\
\hline				                 	                      
Fe\,{\sc ii} & 5500.0 &   3.0 &   5.040 &  4.821 &  (4.777) & +0.020 & (+0.025) & 0.007 \\
Fe\,{\sc ii} & 5500.0 &   3.0 &  30.227 & 29.098 & (27.616) & +0.026 & (+0.062) & 0.010 \\
Fe\,{\sc ii} & 5500.0 &   3.0 &  60.639 & 58.560 & (53.935) & +0.041 & (+0.134) & 0.014 \\
Fe\,{\sc ii} & 5500.0 &   3.0 &  91.386 & 88.180 & (81.324) & +0.051 & (+0.167) & 0.018 \\
\hline				                 	                      
Fe\,{\sc ii} & 4500.0 &   3.0 &   5.179 &  5.011 &  (4.963) & +0.015 & (+0.020) & 0.007 \\
Fe\,{\sc ii} & 5500.0 &   3.0 &   5.040 &  4.821 &  (4.777) & +0.020 & (+0.025) & 0.007 \\
Fe\,{\sc ii} & 6500.0 &   3.0 &   4.947 &  4.689 &  (4.648) & +0.025 & (+0.029) & 0.008 \\
Fe\,{\sc ii} & 7500.0 &   3.0 &   4.878 &  4.591 &  (4.553) & +0.028 & (+0.032) & 0.008 \\
\hline				                 	                      
Fe\,{\sc ii} & 4500.0 &   3.0 &  61.639 & 60.261 & (55.294) & +0.030 & (+0.139) & 0.013 \\
Fe\,{\sc ii} & 5500.0 &   3.0 &  60.639 & 58.560 & (53.935) & +0.041 & (+0.134) & 0.014 \\
Fe\,{\sc ii} & 6500.0 &   3.0 &  59.936 & 57.334 & (52.999) & +0.048 & (+0.130) & 0.014 \\
Fe\,{\sc ii} & 7500.0 &   3.0 &  59.520 & 56.536 & (52.458) & +0.053 & (+0.126) & 0.015 \\ 
\hline\hline
\end{tabular}
\unboldmath
\caption[]
{
Same as Table~\ref{tab1}, but for a sample of fictitious S\,{\sc i}, 
Ca\,{\sc i}, Ca\,{\sc ii}, Ti\,{\sc i}, Ti\,{\sc ii}, Fe\,{\sc i}, 
and Fe\,{\sc ii} lines. 
}
\label{tab3}
\end{center}
\end{table*}

\begin{table*}[htb]
\begin{center}
\boldmath
\begin{tabular}{|r|r|r|r|r r|r r|r|} \hline
{\bfseries Ion} & $\lambda$ [\AA] & $E_{\rm i}$ [eV] & $W_{\rm 1D}$ [m\AA] & 
\multicolumn{2}{c|}{$W_{\rm 2D}$ [m\AA]} &
\multicolumn{2}{c|}{$\Delta_{\rm gran}$} & $\sigma$ \\
\hline\hline

Sr\,{\sc i} &  5500.0 &   0.0 &   7.802 & 12.576 & (12.171) & -0.235 & (-0.218) & 0.098 \\
Sr\,{\sc i} &  5500.0 &   1.0 &   7.233 &  9.569 &  (9.369) & -0.135 & (-0.124) & 0.054 \\
Sr\,{\sc i} &  5500.0 &   2.0 &   6.781 &  7.838 &  (7.701) & -0.069 & (-0.061) & 0.027 \\
Sr\,{\sc i} &  5500.0 &   3.0 &   6.420 &  6.844 &  (6.729) & -0.030 & (-0.022) & 0.012 \\
\hline\hline			                 	                      
				                 	                      
Sr\,{\sc ii} & 5500.0 &   0.0 &   5.552 &  5.740 &  (5.673) & -0.016 & (-0.010) & 0.008 \\
Sr\,{\sc ii} & 5500.0 &   1.0 &   5.369 &  5.299 &  (5.237) & +0.006 & (+0.011) & 0.002 \\
Sr\,{\sc ii} & 5500.0 &   2.0 &   5.236 &  5.069 &  (5.006) & +0.015 & (+0.021) & 0.004 \\
Sr\,{\sc ii} & 5500.0 &   3.0 &   5.153 &  4.972 &  (4.903) & +0.017 & (+0.023) & 0.002 \\
\hline\hline			                 	                      
				                 	                      
Ba\,{\sc ii} & 5500.0 &   0.0 &   5.578 &  5.782 &  (5.678) & -0.017 & (-0.008) & 0.009 \\
Ba\,{\sc ii} & 5500.0 &   1.0 &   5.391 &  5.324 &  (5.229) & +0.006 & (+0.014) & 0.002 \\
Ba\,{\sc ii} & 5500.0 &   2.0 &   5.252 &  5.078 &  (4.982) & +0.016 & (+0.025) & 0.004 \\
Ba\,{\sc ii} & 5500.0 &   3.0 &   5.168 &  4.973 &  (4.869) & +0.018 & (+0.028) & 0.006 \\
\hline\hline

 CN &  5500.0 &   0.0 &   8.865 & 10.974 & (10.877) & -0.100 & (-0.096) & 0.045 \\
 CN &  5500.0 &   1.0 &   8.293 &  9.530 &  (9.465) & -0.065 & (-0.061) & 0.034 \\ 
\hline\hline		                 	                      
			                 	                      
MgH &  5500.0 &   0.0 &   9.744 & 16.181 & (16.004) & -0.246 & (-0.241) & 0.091 \\
MgH &  5500.0 &   1.0 &   9.040 & 12.825 & (12.724) & -0.167 & (-0.163) & 0.061 \\ 
\hline\hline
\end{tabular}
\unboldmath
\caption[]
{
Same as Table~\ref{tab1}, but for a sample of fictitious Sr\,{\sc i}, 
Sr\,{\sc ii}, Ba\,{\sc ii}, CN, and MgH lines. 
}
\label{tab4}
\end{center}
\end{table*}                                                                    
\end{document}